# A Self-Organizing Extreme-Point Tabu-Search Algorithm for Fixed Charge Network Problems with Extensions

Richard S. Barr[1], Fred Glover[2], Toby Huskinson[3], Gary Kochenberger[4]

**Abstract**
We propose a new self-organizing algorithm for fixed-charge network flow problems based on ghost image (GI) processes as proposed in Glover (1994) and adapted to fixed-charge transportation problems in Glover, Amini and Kochenberger (2005). Our self-organizing GI algorithm iteratively modifies an idealized representation of the problem embodied in a parametric ghost image, enabling all steps to be performed with a primal network flow algorithm operating on the parametric GI. Computational tests are carried out on an extensive set of benchmark problems which includes the previous largest set in the literature, comparing our algorithm to the best methods previously proposed for fixed-charge transportation problems, though our algorithm is not specialized to this class. We also provide comparisons for additional more general fixed-charge network flow problems against Cplex 12.8 to demonstrate that the new self-organizing GI algorithm is effective on large problem instances, finding solutions with statistically equivalent objective values at least 700 times faster. The attractive outcomes produced by the current GI/TS implementation provide a significant advance in our ability to solve fixed-cost network problems efficiently and invites its use for larger instances from a variety of application domains.

## 1. Introduction: Problem Definition and Background

We define the network fixed charge problem as

$$\textit{NetFC}: \quad \text{Minimize} \quad x_o[\text{FC}] = cx + F(x)$$
$$\text{Subject to:}$$
$$Ax = b$$
$$U \geq x \geq 0$$

where x is the vector given by x = ($x_j$: j ∈ N = {1, …, n}) and the matrix $A$ is a node-arc incidence matrix, so that the equation $Ax = b$ constitutes a classical network representation of the flow equations defining a pure network problem and the variables $x_j$ correspond to arcs of the network. The fixed charge function is given by $F(x) = \sum(F_j y_j: j \in N)$ where each fixed charge coefficient $F_j$ is nonnegative and the $y_j$ variables take on binary values that satisfy $y_j = 1$ if $x_j > 0$ and $y_j = 0$ otherwise. *F(x)* may be equivalently written as $F(x) = \sum(F_j y_j: j \in N(FC))$, where $N(FC) = \{j \in N: F_j > 0\}$ and we call N(FC) the set of (effective) fixed charge coefficients.

[1] EMIS Department, Lyle School of Engineering, Southern Methodist University, Dallas, TX 75275, barr@smu.edu
[2] ECEE, College of Engineering and Applied Science, University of Colorado – Boulder, Boulder, CO, 80309 USA glover@colorado.edu
[3] Computer Science Department, Lyle School of Engineering, Southern Methodist University, Dallas, TX 75275, thuskins@smu.edu
[4] Business School, University of Colorado at Denver, Denver, CO 80217 USA, gary.kochenberger@ucdenver.edu

Applications of the problem *NetFC* arise in many areas, including facility location, network design, logistics and supply chain, and specific problems, such as lot-sizing, course scheduling, and others. Location problems include the uncapacitated and capacitated facility or plant location problems as described in Fernández and Landete (2015) and Eiselt, Marianov, and Bhadury, (2015). In-depth coverage in Daskin (2013) provides in-depth coverage of the area and extensive list of application papers are available in Eiselt, Marianov, and Bhadury (2015). Network design applications arise in telecommunications (Forsgren and Prytz (2006) and Pioro and Medhi (2004)), including related location problems (Fortz (2015), regional wastewater system design (Jarvis,. Rardin, Unger, Moore and Schimpeler (1978)), and electrical smartgrid data network design, including equipment placement, described in Barr, Jones, and Klinkert (2018).

NetFC problems also have useful applications in supply chain optimization (Alizadeh (2009)), logistics (Alumur, Kara, and Melo (2015)), vanpool assignment (Kaan and Olinick (2013)), and distribution networks (Mateus and Patrocinio (2006)). In addition, they emerge in multi-level lot-sizing within an MRP (Steinberg and H. Albert Napier (1980)) and scheduling training courses (Glover, Klingman, and Phillips (1992)). See other applications enumerated in Nicholson and Zhang, (2016)

In the following, we assume the reader has a basic acquaintance with formulations and solution algorithms for pure networks and is familiar at a rudimentary level with primal simplex algorithms for pure networks. (For references containing useful background information, see for example Ahuja, Magnanti, and Orlin (1993); Bazaraa, Jarvis, and Sherali (2010); and Murty (1992).)

The remainder of this paper is organized as follows. Section 2 introduces our Self-Organizing GI Approach FixNetGI for the network fixed-charge problem and gives a pseudocode for its main algorithm, followed by an explanation of the procedure. Section 3 provides the pseudocodes for the routines invoked by the main algorithm, together with a description of their functions. The design for testing our algorithm and the computational results, together with a comparison involving outcomes obtained by applying the Cplex MIP code [15], are presented in Section 4. As shown, the outcomes demonstrate significant advantages for our algorithm both in solution time and solution quality in solving large and challenging *NetFC* problems. Finally, Section 5 concludes the paper, highlighting the implications of the computational results and identify directions for future research.

## 2. The Self-Organizing GI Approach

The general form of the self-organizing ghost image (GI) approach derives from a collection of problem-solving principles detailed in Glover (1994). Our focus on applying the GI framework to fixed-charge network problems builds on the work of Glover, Amini and Kochenberger (2005) that studies an earlier version of the approach applied to the special case of fixed-charge transportation networks.

Within the pure network setting of *NetFC*, our method exploits the problem structure by introducing a non-negative penalty vector $p = (p_j: j \in N)$ and an associated penalized cost vector given by $c(p) = (c_j + p_j: j \in N)$. The penalty vector $p$ is determined by a self-organizing parameterization to give the following parametric network linear programming relaxation of the fixed-charge problem

**LP(*p*)** Minimize $x_o(p) = c(p)x$
Subject to:

$$Ax = b$$
$$U \geq x \geq 0$$

The parameterization defining $p_j$ occurs by setting $p_j = F_j/v_j$ where $v_j$ denotes a quantity that is systematically updated throughout the algorithm. Hence $p_j$ allocates the fraction $1/v_j$ of the fixed cost $F_j$ to the total cost of $x_j$. We apply the convention that a denominator $v_j$ close to 0 (smaller than a chosen ε value) translates into setting $p_j$ = BigM provided $F_j > 0$, where BigM is a large positive number, and similarly a denominator $v_j$ that exceeds BigM translates into setting $p_j = 0$. However, we will in several instances identify the $p_j$ values directly without bothering to refer to $v_j$. (Note, if $F_j = 0$ then automatically $p_j = 0$ regardless of the value of $v_j$, since $F_j = 0$ expresses the fact that $x_j$ is not a fixed-charge variable. We also interpret the value of $x_j$ to be 0 if this value is less than ε.)

In the case $p = 0$ (where all $p_j = 0$), we have $c(p) = c$, and obtain the simple network linear programming relaxation

**LP:** Minimize $x_o = cx$
Subject to:

$$Ax = b$$
$$U \geq x \geq 0$$

The method sketched in Glover (1994) begins by solving LP, and then solves a succession of problems LP($p$) produced by progressively modifying and updating $p_j$ in alternation with applying an improvement method for enhancing the solution to LP($p$), utilizing adaptive memory strategies from tabu search.

In our adaptation of the self-organizing GI method to the present context, for simplicity we use the convention of identifying the value of the (nonlinear) fixed-charge objective function $x_o$[FC] for a given trial solution vector $x$ (e.g., $x = x', x''$, etc.) as $x_o$ (hence, $x_o' = x_o[\text{FC}:x']$, $x_o'' = x_o[\text{FC}:x'']$, etc.) It is important to keep in mind that in such cases $x_o$ includes reference to the fixed charge component of the objective function, with the sole exception of explicitly referring to the problem LP.

The values $U_o$ and $U_j^o$ defined below are used as *proxy* bounds for $x_j$ that will be introduced to replace the original bound $U_j$ in certain calculations of the algorithm. Apart from trial solution vectors, we maintain a locally optimal solution vector $x^*$ and an overall ("global") best solution vector $x^G$, i.e., $x_o^G$ (= $x_o[\text{FC}:x^G]$) is the minimum of the $x_o^*$ (= $x_o[\text{FC}:x^*]$) values.

We first give a pseudo-code for the main routines of our Ghost Image Tabu Search (GI/TS) method embodied in our FixNetGI code and then describe the rationale that explains the key steps.

## GI/TS Algorithm

The algorithm requires setting the following user input parameters:

*Search limits:*

1. MaxIter: maximum inside loop iterations per invocation
2. MaxPass: number of diversification invocations required to terminate algorithm
3. MaxInsideImprove: number of consecutive non-improving inside loop iterations that will trigger an exit from the inside loop
4. BadLuck: number of consecutive x*-improvement failures that will trigger a diversification
5. OutOfLuck: number of consecutive non-improving outside loop iterations that will trigger an exit from the outside loop.

*Updating $v$:*

6. Alpha(i), i = 1 to 3: weighting factors, summing to 1, for updating $v_j$ values. Weights: Alpha(1) for current $x_j$*, Alpha(2) for current $v_j$ value, and Alpha(3) for the historical $mean_j$ plus $U_j^o$ as adjusted by Beta
7. Beta: weight for historical average associated with Alpha(3) and $v_j$ update
8. MaxSol: when updating $v_j$, the maximum number of previous $x_j$* values used to calculate $mean_j$ for the Alpha(3) term

*Tabu control:*

9. TabuTenure: pivots required before a leaving arc can reenter the network tree (LP basis)

*Duplicate solutions*

10. LimMatch: limits the number of times a solution duplication occurs before triggering diversification
11. sLim: number of solutions saved for duplicate-solution checking
12. ZeroRefresh: number of diversifications performed that will trigger refreshing the duplicate-check solution list with all counts equal 0

The GI/TS algorithm as defined here is supported by several subsidiary procedures to update $v$, control the descent and tabu phases, perform moves/pivots, check for duplicate solutions, and diversity the search. These components are defined and discussed separately.

### *GI/TS Main Routine*

1. **Step 0:** ***solve LP and create initial v, p, and locally best solution x****
   A. Initialize parameters:
      i. JIter= 0, $v_{iter}$ = MaxIter/4, Pass = 0, LastInsideImprove = 0, Zero(s) = (0, … 0) for s = 1 to sLim (i.e., Zero(s,j) = 0 for j = 1 to n), nMatch = 0, Recover = 0, DoTabu = True, NumSol = 0, GbestIter = 0, NoLuck = 0, BigM = large positive number, AscentTenure = DescentTenure = TabuTenure,
      ii. Set $x_o^G$ = BigM
   B. Solve LP, save the solution as the first locally best solution $x^*$ and identify the fixed charge objective function value $x_o^* = x_o[FC{:}x^*]$.
   C. Save the scalar $U_o$ as the largest flow value $x_j$, j ∈ N(FC) in the solution to LP.
   D. Save individual values $U_j^o$ ($\le U_j$) = $x_j$ as the max flow (so far) for each arc j ∈ N(FC).
   E. Set $v_j = U_j$ so that initially $p_j = F_j/U_j$, $mean_j = U_j$ for all j ∈ N(FC).
   **F. Step 1: create and** ***solve LP(p) to get first test solution x'***
      i. Solve LP($p$) by re-optimization to get $x'$ and identify the fixed charge objective function value $x_o' = x_o[FC{:}x']$. NumSol = 1
      ii. Update $U_j^o = \max(U_j^o, x_j')$, for each j ∈ N(FC).
      iii. If $x_o' < x_o^*$ then $x_o^* = x_o'$, $x^* = x'$, set Descent = True and perform V_UPDATE
      iv. Create the $n$-vector ZeroØ, where ZeroØ(j) = 1 if $x_j' = 0$ and $F_j > 0$ (j ∈ N(FC)), else ZeroØ(j) = 0.
      v. Set First = 1 and Zero(1) = SumZeroØ = ZeroØ
      vi. OutsideOK = True
2. **While (OutsideOK):** (Execute the *Outside loop*)
   **A. Step 2:** ***Improve the current solution x', move to local optimum x'', and then to TS improvement***
      *i.* ***Phase I:*** *Refine x' by LP Restriction:*
         a. Set **p**: $p_j$ = BigM if ZeroØ(j) = 1, else $p_j$ = 0
         b. Solve LP(p) by re-optimization to get $x''$ (and $x_o''$)
         c. If $x_o'' < x_o^*$ then set Descent = True (Recording of $x^* = x''$ will be handled later)
         d. If JIter ≤ $v_{iter}$, update $U_j^o = \max(U_j^o, x_j'')$, for each j ∈ N(FC)
      ***ii. Phase II:***

a. Initialize parameters:

  a. Set InsideIter = BestIter = TSImprove = DescentImprove = LastInsideImprove = 0, Descent = True, Improve = False, TabuTenure = DescentTenure

  b. Set Tabu(j) = 0 for each j ∈ N

  c. Set Aspire = Min($x_o''$, $x_o$*). InsideOK = True

b. **Repeat while** (InsideIter < MaxIter and InsideOK) (Execute the *Inside Loop*)

  a. ++InsideIter, j*=k*=0

  b. For every NB arc j ∈ N: Compute $x_{oj}$, the change in the objective function $x_o''$ (= $x_o[FC{:}x'']$) if $x_j$ is pivoted into the basis (and one or more variables $x_k$ are driven to their lower or upper bounds to become candidates to leave the basis). Restrict consideration to j ∈ N satisfying Tabu(j) < InsideIter or satisfying the aspiration criterion $x_{oj}$ < Aspire – $x_o''$

  c. Save the best arc j* = arg min($x_{oj}$: for j subject to the restriction in b.), and identify a leaving arc k*. (k* = j* if there is a "bound flip" where $x_{j^*}$ leaves the basis at its opposite bound)

  d. Perform DESCEND to carry out the pivot and associated update for the choice of j* and k*.

  e. If (InsideIter – LastInsideImprove > MaxInsideImprove) then InsideOK = False (*Exit the Inside Loop*)

  **EndWhile** (for the *Inside Loop*)

c. If ++JIter > MaxIter, OutsideOK = False *(Exit the Outside Loop)*

d. If (Improve) NoLuck = 0

Else

  a. ++NoLuck

  b. If NoLuck = OutOfLuck, OutsideOK = False, BREAK *(Exit the Outside Loop, to conclude at Step 3)*

  ElseIf NoLuck = BadLuck, then

    a. $v_j$ = Max($U_o$ – $v_j$, 1) for each j ∈ N(FC) (mini-diversification)

    b. if $x_o^* < x_o^G$ then update $x_o^G = x_o^*$ and $x^G = x^*$

    c. $x_o^*$ = BigM (to assure LP(p) starts over to make a new local optimum $x^*$)

**B. Create and solve LP(p) to get new test solution x′ and Check for Duplications**

i. Set $p_j = F_j/v_j$ for each j ∈ N(FC)

ii. Solve LP(p) by post-optimization to get $x'$ and $x_o'$

iii. Update $U_o$ = max($U_o$, $x_j'$) for each j ∈ N(FC)

iv. If $x_o' < x_o^*$ then

a. Update $x_o* = x_o'$ and $x* = x'$

b. Perform V_UPDATE

v. Create the n-vector ZeroØ, where ZeroØ(j) = 1 if $F_j > 0$ and $x_j' = 0$, else ZeroØ(j) = 0

vi. Perform DUPCHECK(*which may include DIVERSIFY*)

**Endwhile** (*Outside Loop*)

**3. Conclusion, after exit Outside Loop**

A. If $x_o* < x_o^G$ then $x_o^G = x_o*$ and $x^G = x*$ and set BestPass = Pass

B. STOP

## Discussion of the GI/TS Main Routine

In the initialization step, Step 0, the original linear programming relaxation LP is solved, and its solution is saved as the first locally optimal solution $x*$. Also, to initiate alternative formulas for updating the parameter vector $v$, the constant $U_o$ is initialized to be the largest $x_j$ value obtained in solving LP. In addition, the solution value for each variable $x_j$ is recorded in $U_j^o$.

In Step 1, the problem LP($p$) is solved for the first time by re-optimizing the solution obtained in Step 0 for the modified objective function of LP($p$), to obtain an LP optimum solution, $x'$. The fixed-charge objective function value $x_o'$ (= $x_o$[FC:$x'$]) for $x'$ is calculated and $x'$ replaces the locally best solution $x*$ if $x_o' < x_o*$ (= $x_o$[FC:$x*$]). We continue to update the values $U_j^o$ designated to maintain the maximum value attained by $x_j$ for the first $v_{iter}$ iterations.

To investigate the potential for further improvement to the current solution, $x'$, in Step 2-Phase I the objective function coefficients of the variables with nonzero and zero values are set to their variable costs $c_j$ and $c_j$ + Big$M$, respectively (as a result of setting $p_j = 0$ and $p_j$ = BigM in these two cases), resulting in the specified form of LP($p$), which is then solved by post-optimization, yielding $x''$. The main purpose of setting the cost of variables with the zero values in the trial solution to Big$M$ is to maintain their values at zero during the current post-optimization process, and these variables alternatively could simply be handled by temporarily setting their upper bounds to 0 during this step. Remaining variables that were positive in the solution to the previous LP(p) problem receive their original costs $c_j$ so that the solution will be evaluated relative to the original variable costs. Following the calculation of the fixed charge objective function value for the resulting solution $x''$, the current locally best solution $x*$ is replaced by $x''$ if this new solution turns out to be better. Also, in *Phase I* the value $U_j$, identifying the maximum value for each $x_j$ throughout the first $v_{iter}$ iterations, is updated.

Next, the Inside Loop is initiated within Phase II that executes a tentative pivot exploration process, where each nonbasic variable $x_j$, j ∈ N, is considered as a potential entering variable, and the candidates for the leaving variable, $x_k$, are identified, to determine the change $x_{oj}$ in the fixed charge objective function that would result if $x_j$ were selected to enter the basis tree. The process is guided by a simple tabu search approach, where attention is restricted to j ∈ N satisfying Tabu(j) < InsideIter or satisfying the aspiration criterion $x_{oj}$ < Aspire – $x_o''$, conditions that are irrelevant initially but that become relevant based on updates in the DESCEND routine.

The pseudocodes for the DESCEND procedure and other procedures invoked by the main algorithm appear below, followed by a description of the functions of these procedures.

At the completion of the tentative pivot explorations within the main algorithm, the variable $x_{j*}$ that yields the greatest reduction in the fixed-charge objective function, is selected for pivoting to bring it into the basis. To further improve the current solution, the process returns to the tentative pivot exploration phase, using the current basis representation.

The Inside Loop ends once the current iteration, InsideIter, exceeds the maximum allowed number of iterations, MaxInsideImprove, beyond the last improvement of the locally best solution $x^*$. At the conclusion of the Inside Loop the Outside Loop continues by setting the counter NoLuck to 0 if the Inside Loop had succeeded in improving the locally best solution $x^*$. Otherwise NoLuck is incremented and if NoLuck = OutOfLuck the Outside Loop terminates to record the final global best solution $x^G$ at Step 3. Barring this, if NoLuck = BadLuck, a "mini-diversification" step is initiated. Phase II proceeds to generate the current p vector based on the vector $v$, and then solves LP(p) by post-optimization to obtain $x'$. If the fixed charge objective function value $x_o' = x_o$[FC:$x'$] improves on $x_o^*$ then $x^*$ is updated and the V_UPDATE routine is executed. Finally, the DUPCHECK routine is executed, which may involve executing the DIVERSIFY procedure, to lay the foundation for the next iteration of the Outside Loop.

## 3. Supporting Procedures

We first give the pseudocode for the supporting procedures used within the main routine, in the order in which they first appear in the main routine and in other supporting procedures, and then explain their functions.

### *Procedure DESCEND*

1. If Descent = True then
   - A. If $x_{oj}^*$ < 0 then (*the Decent Phase continues to improve*)
     - i. Perform PIVOTJSTAR (*to pivot in j* and remove k*, …*)
     - ii. Update $x_o''$ and set Aspire = Min($x_o^*$, $x_o''$).
     - iii. ++DescentImprove
   - B. Else
     - i. Descent = False (*happens the first time that leave Descent Phase*)
     - ii. TabuTenure = AscentTenure
     - iii. If $x_o''$ < $x_o^*$ then
       - a. Improve = True
       - b. BestIter = LastInsideImprove = InsideIter – 1 and BestIterG = JIter
       - c. Update $x_o^*$ = $x_o''$ and $x^*$ = $x''$

d. Perform V_UPDATE

*iv*. If DoTabu = False, then InsideOK = False, RETURN (EXIT *Inside loop)*

v. Perform PIVOTJSTAR

2. Else (*Descent = False and we are now in the TS phase*)

A. Perform PIVOTJSTAR

B. If $x_{oj}$* < 0 then

i. TabuTenure = DescentTenure

ii. If $x_o''$ < $x_o$* then

a. Improve = True

b. BestIter = LastInsideImprove = InsideIter, and BestIterG = JIter

c. Update $x_o$* = $x_o''$ and x* = x″

d. ++TSImprove and ++AllTSImprove

e. Aspire = $x_o$*

f. Perform V_UPDATE

C. Else TabuTenure=AscentTenure

3. Update: Tabu(k*) = InsideIter + TabuTenure

4. Return

END DESCEND

### *Procedure V_UPDATE`*

1. ++NumSol
2. Y = Min(NumSol,MaxSol), X = 1/Y
3. For each arc j ∈ N(FC):
   a. $Mean_j = (X)x_j^* + (1\text{-}X)Mean_j$
   b. $UMean = Beta \cdot Mean_j + (1\text{-}Beta)U_o$
   c. $v_j = Alpha(1) \cdot x_j^* + Alpha(2) \cdot v_j + Alpha(3) \cdot UMean$
4. If $x_o^* < x_o^G$ then $x_o^G = x_o^*$ and $x^G = x^*$, GbestIter = JIter

END V_UPDATE

### *Procedure PIVOTJSTAR*

1. Pivot in j* and remove k* from the tree (or perform a bound flip) yielding a new $x''$ and updating $x_o''$.
2. As $x''$ is created, set $U_j^o = \max(U_j^o, x_j'')$ along the basis equivalent path.
3. Return

END PIVOTJSTAR

### *Procedure DUPCHECK*

1. Set s = First and Match = False (*Match will change to True if some Zero(s) = ZeroØ*)
2. For CheckDup = 1 to sLim and Match = False (*DupCheck loop*)
   A. If Zero(CheckDup) = ZeroØ then Match = True, Exit this loop
   B. Else If ++s > sLim then s = 1
3. If Match = True then (*new change*)
   A. If ++nMatch > LimMatch then
      i. sMax = Max(sMax, CheckDup) (*Records how far we had to go to find a match.*)
      ii. Execute DIVERSIFY
      iii. nMatch = 0
4. Else
   A. If nMatch > 0 then
      i. ++Recover
      ii. MaxRecover = Max(Recover,MaxRecover)
      iii. nMatch = 0
   B. SumZeroØ = SumZeroØ + ZeroØ
   C. If First > 1 then Last = First – 1, else Last = sLim.
   D. Replace Zero(Last) by setting Zero(Last) = ZeroØ and set First = Last
5. Return

END DUPCHECK

PROCEDURE DIVERSIFY

1. If $x_o* < x_oG$ then $xG = x^*$ and $x_oG = x_{o*}$ and set BestPass = Pass
2. If Pass = MaxPass then STOP, else ++Pass
3. Let Max = Max(SumZeroØ(j), over j ∈ N(FC))
4. For all j ∈ N(FC)
   A. Let $f_j$ = SumZeroØ(j)/Max
   B. If SumZeroØ(j) > Max/2, $v_j = \lceil f_j \cdot U_j \rceil$
   C. Else $v_j = \text{Max}(\lceil f_j \cdot U_j^o \rceil, 1)$
   *D.* $p_j = F_j/v_j$
5. Create and solve LP(p) to get new “first” test solution x’
   ***A.*** Solve LP(p) by post-optimization to get x′ and $x_o'$
   B. Begin x* again from scratch to set x* = x′ and $x_o^* = x_o'$
   C. Update $U_j^o = \max(U_j^o, x_j')$ for each j ∈ N(FC)
   D. Create the n-vector ZeroØ, where ZeroØ(j) = 1 if $F_j > 0$ and $x_j' = 0$, else ZeroØ(j) = 0

E. Perform V_UPDATE

6. Set First = 1 and Zero(1) = ZeroØ
7. Set Zero(s) = (0, … 0) for s = 2 to sLim
8. If Pass is a multiple of ZeroRefresh, then also re-initialize SumZeroØ = (0, … 0), but otherwise let SumZeroØ continue to accumulate

END DIVERSIFY

## Discussion of the Supporting Procedures

The DESCEND routine is the first supporting procedure invoked by the main routine, to implement the choice of $x_{j*}$ as the incoming pivot variable and the associated $x_{k*}$ as the leaving variable. If the algorithm is in a descent phase (Descent = True and TabuTenure = DescentTenure), and if the value $x_{oj*}$ continues the descent ($x_{oj*} < 0$), then the routine simply performs the PIVOTJSTAR procedure which pivots in $x_{j*}$ and removes $x_{k*}$ from the basis tree, to produce the updated solution x" and its fixed charge objective $x_o$", and updates $U_j^o$ for variables along the basis exchange path. Once the descent ends, Descent is set to False, TabuTenure is set to AscentTenure, and a check is performed to see if the solution x" (before updating by the basis exchange of $x_{j*}$ and $x_{k*}$) improves on $x^*$ ($x_o'' < x_o^*$). In this case, $x^*$ is updated as customary and the routine performs V_UPDATE, which updates the $v_j$ values as a foundation for subsequently determining the $p_j$ values that define the problem LP($p$). PIVOTJSTAR is likewise performed now that the descent ends.

When the DESCEND routine is invoked and Descent = False, the PIVOTJSTAR routine is immediately performed and if $x_o'' < x_o^*$, then x* is updated as before. (The value $x_{oj*}$ can be improving after the initial descent has concluded. Instead of bouncing in and out of successive descent and ascent phases, once the initial descent has concluded, all subsequent steps are treated as an "ascent tabu phase." However, TabuTenure is set to DescentTenure whenever an improving step occurs, and to AssetTenure otherwise.) Finally, Tabu(k*) = InsideIter + TabuTenure for the variable $x_{k*}$ that leaves the basis tree and becomes non-basic.

Having discussed V_CHECK and PIVOTJSTAR in the explanation of DESCEND, it remains to discuss the supporting procedure DUPCHECK and the DIVERSIFY procedure that is invoked within it.

The DUPCHECK routine is designed to check whether there are any duplications among the most recent ZeroØ vectors stored in Zero(s) for s = 1 to sLim. Since each ZeroØ vector identifies the variables $x_j$ that equal 0 in a given solution (by setting ZeroØ(j) = 1), and setting these variables to 0 automatically determines the network solution that sets remaining variables to 1, a duplication in these vectors implies that the associated fixed charge solutions are duplicated. DUPCHECK carries out a check for duplications (matches) by recording Zero(s) as a wraparound list, where the most recent ZeroØ vector is stored in Zero(First) and Zero(Last) is the ZeroØ vector recorded sLim iterations ago. The Zero(s) array starts from s= First until reaching s = sLim, and then continues at s = 1 until reaching s = First – 1. Then the new (now most recent) ZeroØ vector is recorded by writing over the oldest one in the location s = First – 1 and then First is updated by setting First = First – 1. (Special case: If First = 1 then the location First – 1 is sLim.) This device avoids having to write the vectors into a temporary array and then write them back into Zero(s) to allow Zero(s) to always go from s = 1 to sLim.

If the number of matches nMatch is found to exceed the limit LimMatch, the DIVERSIFY routine is executed which updates $x^G$ if the current x* improves upon it and if the DIVERSIFY routine has been invoked MaxPass times the algorithm stops. Otherwise the diversification proceeds by generating new $f_j$ values based on the formula $f_j$ = SumZeroØ(j)/Max, where SumZeroØ(j) counts the number of times $x_j = 0$ in a solution that produced a ZeroØ vector in the DUPCHECK routine, and Max is the maximum of these SumZeroØ(j) values. The new $v_j$ values are then determined by setting $v_j = \lceil f_j \cdot U_j \rceil$ if SumZeroØ(j) > Max/2 and otherwise setting $v_j$ = Max($\lceil f_j \cdot U_j^o \rceil$ , 1).

From this, the $p_j$ values are determined by the usual formula $p_j = F_j/v_j$ as a basis for creating the problem LP(p) which is then solved by post-optimization to obtain a solution x'. The locally optimal solution x* starts again "from scratch" by setting x* = x', and the bounds $U_j^o$ are updated in the customary way, along with establishing the ZeroØ vector as in the first step of the main algorithm. Finally, the V_UPDATE routine is executed, and the arrays associated with ZeroØ are likewise re-initialized, to conclude the DIVERSIFY procedure.

In the event than Match is not True in the DUPCHECK procedure (and hence nMatch is not checked for exceeding LimMatch, and DIVERSIFY is not executed), then the DUPCHECK procedure updates values for tracking the algorithm's performance, assures that nMatch = 0, and updates the Zero(s) array in accordance with the explanation above.

In conjunction with the main routine, these supporting procedures complete the GI/TS algorithm.

## 4. GI/TS Computational Testing

An implementation of the above GI/TS algorithm, our code FixNetGI, was built using the alternating–path primal network simplex methods and data structures described in [1, 2, 3]. This solver is implemented in Fortran, compiled with `gfortran -O3`, and tested under the Centos 6.10 version of the Linux operating system at Southern Methodist University. The test hardware is a Dell R720 with a Dual Six Core Intel Xeon @ 3.5GHz with all runs executed in single-thread mode.

To assess the performance of FixNetGI, computational comparisons in terms of solution quality and speed are made with the IBM commercial optimization software Cplex 12.8, running with default parameters except for specifying single-threaded execution mode and a time limit per problem. Since Cplex is a general-purpose optimizer for linear and mixed-integer problems, the special-purpose heuristic approach of FixNetGI gives it major advantages. This comparison, however, is valuable because: no comparable solver for NetFC is available, Cplex is widely used and respected by practitioners and researchers, and the comparison will indicate the heuristic's efficiency and solution quality for use on real-world industry problems of this type.

To test the effectiveness of the new solution approach, two problem test sets are used for benchmarking. The first is a collection of known problems from the literature and the second is a new suite of larger problems generated to explore the effects of problem characteristics on performance.

Since there are over a dozen tuning parameters for the heuristic, we performed preliminary testing to identify a single set of parameters to use for all computational results reported herein. Randomly selected values from assigned ranges were run on the test sets, giving quite varied results, but providing guidance as to what value ranges seemed appropriate. The following parameter settings are employed for all runs reported: MaxIter = 50, MaxPass = 10, MaxInsideImprove = 40, BadLuck = 5, OutOfLuck = 20, Alpha(1) = 0.3, Alpha(2) = 0.45, Alpha(3) = 0.25, Beta = 0.4, MaxSolLimit = 1000, TabuTenure = 10, LimMatch = 10, sLim = 10, and ZeroRefresh = 30.

## Test Set 1: Description

This first set of studied problems is drawn from the comprehensive FCTP testbed of Sun, et al [4] with a variety of problem dimensions and characteristics. The problems were originally created with a version of the well-known NETGEN random problem generator [6, 7], modified to include fixed costs on arcs.

These Test Set 1 problems have seven problem dimensions, eight fixed-cost ranges (or types, labeled A-H), and 17 randomly generated instances of each combination. See Table 1 for definitions of these characteristics.

**Table 1. Test Set 1 problem characteristics: (a) dimensions, (b) fixed cost range [4]**

| Problem Dimensions | Total Supply |
|---|---|
| 10 X 10 | 10,000 |
| 10 X 20 | 15,000 |
| 15 X 15 | 15,000 |
| 10 X 30 | 15,000 |
| 50 X 50 | 50,000 |
| 30 X 100 | 30,000 |
| 50 X 100 | 50,000 |

(a)

| Fixed-charge Type | Fixed-charge Range |
|---|---|
| A | [50, 200] |
| B | [100, 400] |
| C | [200, 800] |
| D | [400, 1,600] |
| E | [800, 3,200] |
| F | [1,600, 6,400] |
| G | [3,200, 12,800] |
| H | [6,400, 25,600] |

(b)

Each test problem is a totally dense capacitated fixed-charge transportation problem with randomly distributed supplies and demands per Table 1(a) and with each arc randomly assigned a discrete variable cost between 3 and 8 plus a fixed cost in the associated range from Table 1(b).

A subset of the 896 original testbed problems were selected for computational experiments with the GI2 code, following the choices of [5]. For the six smallest problem sizes, two instances of type A were used for this experimentation. For the largest and most difficult 50x100 size, all 15 instances of each fixed-charge type (A-H) were included, for a total of 132 problems. Hence the focus is on mixed-integer programs with 50,000 binary variables.

## Test Set 1: Computational Results and Analysis

Table 2 describes the solution results for the 12 smaller problems tested. Shown are the dimensions of the transportation problem, the problem identifier, the best solution value found and CPU solution time for Cplex 12.8 (run with a 7200-second time limit) and the FixNetGI code, the ratio of the two solvers' solution values (Z-ratio = FixNetGI's $x_o^G$ /CPLEX $z^*$) and the Cplex time as a multiple of the FixNetGI solution time (Time-X).

Table 2 **Test Set 1** solution results for small problems, type A

| | | CPLEX 12.8 | | FixNetGI | | | |
|---|---|---|---|---|---|---|---|
| Dimension | Prob ID | Best Z | Time (sec) | Best Z | Time (sec) | Z-Ratio | Time-X |
| 10x10 | N104 | 40,255 | 1.49 | 40,258 | 0.01 | 1.0001 | 114.62 |
| 10x10 | N107 | 42,026 | 1.16 | 42,029 | 0.01 | 1.0001 | 116.00 |
| 10x20 | N304 | 56,361 | 0.74 | 56,366 | 0.02 | 1.0001 | 32.17 |
| 10x20 | N307 | 49,737 | 1.61 | 49,742 | 0.03 | 1.0001 | 59.63 |
| 15x15 | N204 | 54,497 | 1.48 | 54,547 | 0.03 | 1.0009 | 49.33 |
| 15x15 | N207 | 53,591 | 1.26 | 53,601 | 0.03 | 1.0002 | 43.45 |
| 10x30 | N504 | 56,883 | 3.20 | 57,137 | 0.04 | 1.0045 | 78.05 |
| 10x30 | N507 | 52,898 | 4.72 | 52,998 | 0.04 | 1.0019 | 134.86 |
| 50x50 | N1004 | 162,863 | 7,200.03 | 163,764 | 1.64 | 1.0055 | 4,395.62 |
| 50x50 | N1007 | 161,186 | 7,200.00 | 162,386 | 0.56 | 1.0074 | 12,834.22 |
| 30x100 | N2004 | 103,163 | 7,200.00 | 104,204 | 0.57 | 1.0101 | 12,543.55 |
| 30x100 | N2007 | 103,402 | 7,200.00 | 104,340 | 0.55 | 1.0091 | 13,162.71 |
| Average: | | 78,072 | 2,401.31 | 78,448 | 0.29 | 1.0033 | 3,630.35 |

With these smaller problems, the heuristic's $x_o^G$ solution values are within 0.1% of the Cplex optimal, on average, and were identified an average of three orders of magnitude faster. One third of FixNetGI's solutions were optimal and its solution times averaged half a second.

The bulk of the testing was focused on the more-difficult totally dense fixed-charge transportation problems with 50 source and 100 sink nodes, 50,000 arcs, supply of 50,000, and all fixed charge ranges as described in Table 1(b). Table 3 summarizes the results from solving 15 problem instances from each of the eight fixed-charge ranges (A-H). Detailed computational results from these 120 problems are found in Tables 4-11.

The results on the larger problems underscore the effectiveness of the GI/TS algorithm. In every case, Cplex did not run to completion and exited at the 7,200-second time limit, while FixNetGI used an average of 1.11 seconds of CPU time. Although FixNetGI's solution values averaged 9% higher, these were identified 6,000 times faster.

To evaluate these solvers' ability to handle even more challenging problems, such as is found in industrial applications, a new problem set was created. The problems are not only larger, but the suite is structured to facilitate statistical analysis of problem characteristics.

**Table 3. Test Set 1: summary of difficult, large 50x100 problems, averages of 15 problems per fixed-charge type**

| Fixed-Charge | | CPLEX 12.8 | | FixNetGI | | | |
|---|---|---|---|---|---|---|---|
| Type | Range | Best Z | Time (sec) | Best Z | Time (sec) | Z-Ratio | Time-X |
| A | 50-200 | 165,809 | 7,200.01 | 167,499 | 1.09 | 1.010 | 6,589 |
| B | 100-400 | 175,337 | 7,200.00 | 178,795 | 1.09 | 1.020 | 6,614 |
| C | 200-800 | 193,422 | 7,200.00 | 200,498 | 1.22 | 1.037 | 5,917 |
| D | 400-1600 | 227,260 | 7,200.00 | 241,310 | 1.09 | 1.062 | 6,625 |
| E | 800-3200 | 289,470 | 7,200.01 | 316,637 | 1.08 | 1.094 | 6,675 |
| F | 1600-6400 | 405,351 | 7,200.00 | 459,073 | 1.08 | 1.133 | 6,674 |
| G | 3200-12800 | 624,726 | 7,200.00 | 731,128 | 1.19 | 1.170 | 6,303 |
| H | 6400-25600 | 1,046,011 | 7,200.01 | 1,258,395 | 1.08 | 1.203 | 6,664 |
| Average: | | 390,923 | 7,200.01 | 444,167 | 1.11 | 1.091 | 6,508 |

**Table 4. Test Set 1: solution results for larger, difficult problems, type A fixed costs in range [50, 200]**

| | | Prob Type | CPLEX 12.8 | | FixNetGI | | | |
|---|---|---|---|---|---|---|---|---|
| PROB | Size | FC range | Best Z | Time (sec) | Best Z | Time (sec) | Z-Ratio | Time-X |
| N3001 | 50x100 | A | 165,214 | 7,200.00 | 166,974 | 1.12 | 1.011 | 6,446 |
| N3002 | 50x100 | A | 166,266 | 7,200.01 | 168,050 | 1.11 | 1.011 | 6,516 |
| N3003 | 50x100 | A | 167,095 | 7,200.00 | 168,503 | 1.16 | 1.008 | 6,212 |
| N3004 | 50x100 | A | 165,793 | 7,200.00 | 167,406 | 1.11 | 1.010 | 6,516 |
| N3005 | 50x100 | A | 166,360 | 7,200.00 | 168,106 | 1.07 | 1.010 | 6,754 |
| N3006 | 50x100 | A | 164,614 | 7,200.01 | 166,146 | 1.06 | 1.009 | 6,818 |
| N3007 | 50x100 | A | 166,007 | 7,200.02 | 167,552 | 1.11 | 1.009 | 6,516 |
| N3008 | 50x100 | A | 164,273 | 7,200.02 | 165,943 | 1.09 | 1.010 | 6,618 |
| N3009 | 50x100 | A | 165,641 | 7,200.01 | 167,421 | 1.07 | 1.011 | 6,761 |
| N300A | 50x100 | A | 166,124 | 7,200.03 | 167,635 | 1.04 | 1.009 | 6,930 |
| N300B | 50x100 | A | 167,103 | 7,200.02 | 168,913 | 1.05 | 1.011 | 6,870 |
| N300C | 50x100 | A | 163,857 | 7,200.00 | 165,929 | 1.09 | 1.013 | 6,624 |
| N300D | 50x100 | A | 164,909 | 7,200.01 | 166,494 | 1.10 | 1.010 | 6,534 |
| N300E | 50x100 | A | 168,075 | 7,200.00 | 169,908 | 1.17 | 1.011 | 6,138 |
| | Average: | | 165,809 | 7,200.01 | 167,499 | 1.09 | 1.010 | 6,589 |

**Table 5. Test Set 1: solution results for larger, difficult problems, type B fixed costs in range [100, 400]**

| PROB | Size | Prob Type FC range | CPLEX 12.8 Best Z | CPLEX 12.8 Time (sec) | FixNetGI Best Z | FixNetGI Time (sec) | Z-Ratio | Time-X |
|---|---|---|---|---|---|---|---|---|
| N3100 | 50x100 | B | 176,223 | 7,200.00 | 179,323 | 1.04 | 1.018 | 6,943 |
| N3101 | 50x100 | B | 174,779 | 7,200.00 | 178,546 | 1.07 | 1.022 | 6,704 |
| N3102 | 50x100 | B | 175,859 | 7,200.00 | 179,340 | 1.08 | 1.020 | 6,667 |
| N3103 | 50x100 | B | 176,296 | 7,200.00 | 179,287 | 1.15 | 1.017 | 6,245 |
| N3104 | 50x100 | B | 176,175 | 7,200.01 | 179,947 | 1.15 | 1.021 | 6,283 |
| N3105 | 50x100 | B | 175,673 | 7,200.00 | 179,081 | 1.06 | 1.019 | 6,825 |
| N3106 | 50x100 | B | 174,171 | 7,200.01 | 177,536 | 1.10 | 1.019 | 6,545 |
| N3107 | 50x100 | B | 175,253 | 7,200.01 | 178,562 | 1.13 | 1.019 | 6,400 |
| N3108 | 50x100 | B | 173,440 | 7,200.01 | 177,091 | 1.12 | 1.021 | 6,457 |
| N3109 | 50x100 | B | 174,661 | 7,200.00 | 178,350 | 1.06 | 1.021 | 6,825 |
| N310A | 50x100 | B | 176,295 | 7,200.01 | 179,663 | 1.08 | 1.019 | 6,691 |
| N310B | 50x100 | B | 176,731 | 7,200.00 | 180,122 | 1.06 | 1.019 | 6,825 |
| N310C | 50x100 | B | 173,012 | 7,200.01 | 176,917 | 1.08 | 1.023 | 6,698 |
| N310D | 50x100 | B | 174,555 | 7,200.00 | 177,726 | 1.07 | 1.018 | 6,748 |
| N310E | 50x100 | B | 176,933 | 7,200.01 | 180,436 | 1.13 | 1.020 | 6,349 |
| | Average: | | 175,274 | 7,200.01 | 178,757 | 1.09 | 1.020 | 6,590 |

**Table 6. Test Set 1: solution results for larger, difficult problems, type C fixed costs in range [200, 800]**

| PROB | Size | Prob Type FC range | CPLEX 12.8 Best Z | CPLEX 12.8 Time (sec) | FixNetGI Best Z | FixNetGI Time (sec) | Z-Ratio | Time-X |
|---|---|---|---|---|---|---|---|---|
| N3200 | 50x100 | C | 194,225 | 7,200.00 | 201,498 | 1.05 | 1.037 | 6,844 |
| N3201 | 50x100 | C | 193,288 | 7,200.00 | 200,823 | 1.12 | 1.039 | 6,440 |
| N3202 | 50x100 | C | 194,189 | 7,200.00 | 202,126 | 1.08 | 1.041 | 6,660 |
| N3203 | 50x100 | C | 193,755 | 7,200.00 | 200,250 | 1.12 | 1.034 | 6,434 |
| N3204 | 50x100 | C | 195,218 | 7,200.01 | 202,696 | 1.12 | 1.038 | 6,406 |
| N3205 | 50x100 | C | 193,750 | 7,200.00 | 200,086 | 1.06 | 1.033 | 6,805 |
| N3206 | 50x100 | C | 192,095 | 7,200.00 | 199,228 | 1.08 | 1.037 | 6,679 |
| N3207 | 50x100 | C | 192,863 | 7,200.00 | 199,989 | 1.06 | 1.037 | 6,786 |
| N3208 | 50x100 | C | 191,262 | 7,200.01 | 197,823 | 1.10 | 1.034 | 6,545 |
| N3209 | 50x100 | C | 192,371 | 7,200.00 | 199,614 | 1.06 | 1.038 | 6,773 |
| N320A | 50x100 | C | 195,345 | 7,200.02 | 201,847 | 1.08 | 1.033 | 6,679 |
| N320B | 50x100 | C | 195,428 | 7,200.00 | 202,049 | 1.06 | 1.034 | 6,786 |
| N320C | 50x100 | C | 190,533 | 7,200.01 | 197,625 | 1.98 | 1.037 | 3,640 |
| N320D | 50x100 | C | 192,668 | 7,200.00 | 199,815 | 2.15 | 1.037 | 3,347 |
| N320E | 50x100 | C | 194,341 | 7,200.00 | 202,005 | 1.13 | 1.039 | 6,377 |
| | Average: | | 193,365 | 7,200.00 | 200,427 | 1.23 | 1.037 | 6,169 |

**Table 7. Test Set 1: solution results for larger, difficult problems, type D fixed costs in range [400, 1600]**

| | | Prob Type | CPLEX 12.8 | | FixNetGI | | | |
|---|---|---|---|---|---|---|---|---|
| PROB | Size | FC range | Best Z | Time (sec) | Best Z | Time (sec) | Z-Ratio | Time-X |
| N3300 | 50x100 | D | 228,374 | 7,200.01 | 241,643 | 1.05 | 1.058 | 6,857 |
| N3301 | 50x100 | D | 227,575 | 7,200.01 | 242,211 | 1.11 | 1.064 | 6,498 |
| N3302 | 50x100 | D | 228,110 | 7,200.01 | 242,684 | 1.10 | 1.064 | 6,575 |
| N3303 | 50x100 | D | 225,815 | 7,200.00 | 239,882 | 1.12 | 1.062 | 6,440 |
| N3304 | 50x100 | D | 229,561 | 7,200.00 | 244,426 | 1.16 | 1.065 | 6,218 |
| N3305 | 50x100 | D | 227,701 | 7,200.01 | 241,937 | 1.07 | 1.063 | 6,710 |
| N3306 | 50x100 | D | 226,219 | 7,200.01 | 239,843 | 1.08 | 1.060 | 6,679 |
| N3307 | 50x100 | D | 225,348 | 7,200.00 | 239,331 | 1.09 | 1.062 | 6,636 |
| N3308 | 50x100 | D | 224,414 | 7,200.00 | 236,798 | 1.10 | 1.055 | 6,551 |
| N3309 | 50x100 | D | 226,652 | 7,200.00 | 241,535 | 1.09 | 1.066 | 6,630 |
| N330A | 50x100 | D | 231,382 | 7,200.00 | 244,641 | 1.05 | 1.057 | 6,844 |
| N330B | 50x100 | D | 230,094 | 7,200.00 | 244,703 | 1.04 | 1.063 | 6,916 |
| N330C | 50x100 | D | 224,210 | 7,200.01 | 238,289 | 1.06 | 1.063 | 6,812 |
| N330D | 50x100 | D | 226,083 | 7,200.00 | 241,055 | 1.07 | 1.066 | 6,729 |
| N330E | 50x100 | D | 227,364 | 7,200.00 | 240,667 | 1.13 | 1.059 | 6,360 |
| | Average: | | 227,181 | 7,200.00 | 241,286 | 1.09 | 1.062 | 6,614 |

**Table 8. Test Set 1: solution results for larger, difficult problems, type E fixed costs in range [800, 3200]**

| | | Prob Type | CPLEX 12.8 | | FixNetGI | | | |
|---|---|---|---|---|---|---|---|---|
| PROB | Size | FC range | Best Z | Time (sec) | Best Z | Time (sec) | Z-Ratio | Time-X |
| N3400 | 50x100 | E | 291,035 | 7,200.00 | 316,495 | 1.03 | 1.087 | 6,970 |
| N3401 | 50x100 | E | 289,261 | 7,200.01 | 316,734 | 1.07 | 1.095 | 6,754 |
| N3402 | 50x100 | E | 290,616 | 7,200.01 | 319,367 | 1.08 | 1.099 | 6,667 |
| N3403 | 50x100 | E | 284,639 | 7,200.00 | 310,945 | 1.12 | 1.092 | 6,434 |
| N3404 | 50x100 | E | 292,426 | 7,200.04 | 321,563 | 1.13 | 1.100 | 6,389 |
| N3405 | 50x100 | E | 290,940 | 7,200.00 | 318,012 | 1.05 | 1.093 | 6,870 |
| N3406 | 50x100 | E | 288,448 | 7,200.00 | 314,592 | 1.08 | 1.091 | 6,698 |
| N3407 | 50x100 | E | 284,681 | 7,200.00 | 311,924 | 1.09 | 1.096 | 6,599 |
| N3408 | 50x100 | E | 285,990 | 7,200.00 | 312,083 | 1.08 | 1.091 | 6,654 |
| N3409 | 50x100 | E | 289,127 | 7,200.01 | 316,983 | 1.07 | 1.096 | 6,710 |
| N340A | 50x100 | E | 296,495 | 7,200.01 | 324,751 | 1.07 | 1.095 | 6,754 |
| N340B | 50x100 | E | 293,248 | 7,200.00 | 320,955 | 1.06 | 1.094 | 6,786 |
| N340C | 50x100 | E | 287,021 | 7,200.01 | 315,303 | 1.07 | 1.099 | 6,716 |
| N340D | 50x100 | E | 288,295 | 7,200.01 | 313,506 | 1.08 | 1.087 | 6,661 |
| N340E | 50x100 | E | 289,837 | 7,200.00 | 316,340 | 1.12 | 1.091 | 6,457 |
| | Average: | | 289,359 | 7,200.01 | 316,647 | 1.08 | 1.094 | 6,654 |

**Table 9. Test Set 1: solution results for larger, difficult problems, type F fixed costs in range [1600, 6400]**

| | | Prob Type | CPLEX 12.8 | | FixNetGI | | | |
|---|---|---|---|---|---|---|---|---|
| PROB | Size | FC range | Best Z | Time (sec) | Best Z | Time (sec) | Z-Ratio | Time-X |
| N3500 | 50x100 | F | 406,610 | 7,200.00 | 462,061 | 1.05 | 1.136 | 6,890 |
| N3501 | 50x100 | F | 403,755 | 7,200.00 | 460,160 | 1.08 | 1.140 | 6,667 |
| N3502 | 50x100 | F | 405,202 | 7,200.00 | 459,936 | 1.09 | 1.135 | 6,581 |
| N3503 | 50x100 | F | 394,992 | 7,200.01 | 445,519 | 1.11 | 1.128 | 6,475 |
| N3504 | 50x100 | F | 409,471 | 7,200.01 | 464,457 | 1.09 | 1.134 | 6,630 |
| N3505 | 50x100 | F | 407,823 | 7,200.00 | 462,557 | 1.04 | 1.134 | 6,923 |
| N3506 | 50x100 | F | 403,233 | 7,200.00 | 450,885 | 1.07 | 1.118 | 6,704 |
| N3507 | 50x100 | F | 396,770 | 7,200.00 | 452,211 | 1.10 | 1.140 | 6,534 |
| N3508 | 50x100 | F | 402,621 | 7,200.00 | 457,526 | 1.09 | 1.136 | 6,606 |
| N3509 | 50x100 | F | 405,749 | 7,200.00 | 460,973 | 1.08 | 1.136 | 6,642 |
| N350A | 50x100 | F | 415,374 | 7,200.00 | 464,597 | 1.06 | 1.119 | 6,792 |
| N350B | 50x100 | F | 409,530 | 7,200.00 | 462,858 | 1.10 | 1.130 | 6,575 |
| N350C | 50x100 | F | 405,979 | 7,200.00 | 459,980 | 1.07 | 1.133 | 6,729 |
| N350D | 50x100 | F | 405,994 | 7,200.00 | 459,980 | 1.07 | 1.133 | 6,735 |
| N350E | 50x100 | F | 407,160 | 7,200.00 | 462,399 | 1.09 | 1.136 | 6,624 |
| | Average: | | 405,261 | 7,200.00 | 458,860 | 1.08 | 1.132 | 6,658 |

**Table 10. Test Set 1: solution results for larger, difficult problems, type G fixed costs in range [3200, 12800]**

| | | Prob Type | CPLEX 12.8 | | FixNetGI | | | |
|---|---|---|---|---|---|---|---|---|
| PROB | Size | FC range | Best Z | Time (sec) | Best Z | Time (sec) | Z-Ratio | Time-X |
| N3600 | 50x100 | G | 628,353 | 7,200.00 | 728,685 | 1.05 | 1.160 | 6,851 |
| N3601 | 50x100 | G | 623,633 | 7,200.00 | 728,390 | 1.07 | 1.168 | 6,748 |
| N3602 | 50x100 | G | 622,435 | 7,200.01 | 739,308 | 1.07 | 1.188 | 6,742 |
| N3603 | 50x100 | G | 606,551 | 7,200.02 | 706,872 | 1.11 | 1.165 | 6,463 |
| N3604 | 50x100 | G | 629,427 | 7,200.00 | 733,056 | 1.83 | 1.165 | 3,934 |
| N3605 | 50x100 | G | 627,022 | 7,200.00 | 729,120 | 2.07 | 1.163 | 3,483 |
| N3606 | 50x100 | G | 623,664 | 7,200.00 | 726,111 | 1.10 | 1.164 | 6,569 |
| N3607 | 50x100 | G | 609,916 | 7,200.00 | 718,671 | 1.11 | 1.178 | 6,516 |
| N3608 | 50x100 | G | 621,534 | 7,200.01 | 724,269 | 1.09 | 1.165 | 6,630 |
| N3609 | 50x100 | G | 623,355 | 7,200.00 | 738,275 | 1.06 | 1.184 | 6,792 |
| N360A | 50x100 | G | 638,942 | 7,200.00 | 735,655 | 1.07 | 1.151 | 6,735 |
| N360B | 50x100 | G | 632,751 | 7,200.00 | 744,229 | 1.07 | 1.176 | 6,761 |
| N360C | 50x100 | G | 627,701 | 7,200.00 | 741,241 | 1.06 | 1.181 | 6,812 |
| N360D | 50x100 | G | 627,689 | 7,200.00 | 741,241 | 1.06 | 1.181 | 6,805 |
| N360E | 50x100 | G | 627,919 | 7,200.00 | 731,792 | 1.08 | 1.165 | 6,698 |
| | Average: | | 624,467 | 7,200.00 | 731,302 | 1.20 | 1.171 | 6,263 |

**Table 11. Test Set 1: solution results for larger, difficult problems, type H fixed costs in range [6400, 25600]**

| PROB | Size | Prob Type FC range | CPLEX 12.8 Best Z | CPLEX 12.8 Time (sec) | FixNetGI Best Z | FixNetGI Time (sec) | Z-Ratio | Time-X |
|---|---|---|---|---|---|---|---|---|
| N3700 | 50x100 | H | 1,054,655 | 7,200.00 | 1,266,006 | 1.07 | 1.200 | 6,754 |
| N3701 | 50x100 | H | 1,041,146 | 7,200.00 | 1,263,578 | 1.07 | 1.214 | 6,704 |
| N3702 | 50x100 | H | 1,040,325 | 7,200.01 | 1,252,861 | 1.09 | 1.204 | 6,636 |
| N3703 | 50x100 | H | 1,018,972 | 7,200.02 | 1,239,035 | 1.10 | 1.216 | 6,522 |
| N3704 | 50x100 | H | 1,050,443 | 7,200.00 | 1,263,694 | 1.09 | 1.203 | 6,593 |
| N3705 | 50x100 | H | 1,053,995 | 7,200.01 | 1,263,791 | 1.07 | 1.199 | 6,704 |
| N3706 | 50x100 | H | 1,049,237 | 7,200.01 | 1,260,282 | 1.08 | 1.201 | 6,661 |
| N3707 | 50x100 | H | 1,022,451 | 7,200.01 | 1,229,135 | 1.10 | 1.202 | 6,563 |
| N3708 | 50x100 | H | 1,040,737 | 7,200.01 | 1,255,743 | 1.10 | 1.207 | 6,528 |
| N3709 | 50x100 | H | 1,041,100 | 7,200.01 | 1,255,976 | 1.08 | 1.206 | 6,667 |
| N370A | 50x100 | H | 1,067,181 | 7,200.01 | 1,281,905 | 1.08 | 1.201 | 6,685 |
| N370B | 50x100 | H | 1,061,167 | 7,200.01 | 1,280,281 | 1.08 | 1.206 | 6,685 |
| N370C | 50x100 | H | 1,052,506 | 7,200.01 | 1,260,941 | 1.07 | 1.198 | 6,761 |
| N370D | 50x100 | H | 1,052,254 | 7,200.00 | 1,260,941 | 1.07 | 1.198 | 6,761 |
| N370E | 50x100 | H | 1,044,003 | 7,200.02 | 1,241,756 | 1.07 | 1.189 | 6,735 |
| | Average: | | 1,045,394 | 7,200.01 | 1,257,851 | 1.08 | 1.203 | 6,657 |

## Test Set 2: Overview and Experimental design

To explore still larger problems and the possible effects of problem structure on solution time and quality, an experimental design using randomly generated test problems was established. For this, the NETGEN problem generator [9], modified to include fixed charges, created a new structured suite of transportation and transshipment problems with up to 33 times as many nodes, 100,000 binary variables, and a variety of problem characteristics.

Test Set 2 consists of 96 problems, each generated with a different seed value, and with problem characteristics varied to enable a full-factorial experimental design. All combinations of five factors are used: number of problem nodes (500, 1000, 3000, and 5000), percentage of source and sink nodes (30% / 70%, transportation, and 20% / 20%, transshipment), number of arcs (10,000, 50,000, and 100,000), total supply (100,000 and 500,000), and fixed-cost range (20-200 and 1600-6400). All arcs have a fixed cost, a variable cost between 3-8, and an arc capacity between 200 and 1500 units. Transshipment sources and sinks are not used.

Tables 12 and 13 display Test Set 2's problem characteristics and solution results from the FixNetGI code and Cplex 12.8, run with a one-hour time limit and a single CPU thread. Problem characteristics shown are problem identifier and the number of nodes, sources and sinks, arcs, total supply, and fixed-cost range. Solution results are: the best solution value found (Best Z) for each application, the ratio of FixNetGI Z to Cplex Z (Z-ratio), the solution time using FixNet, and the Cplex time (3600 seconds in all instances) as a multiple of the FixNet solution time (Cplex Time-X).

**Table 12 Test Set 2, 500- and 1000-node problem characteristics and solution results for FixNetGI and Cplex 12.8**

| Prob | Nodes | Sources / Sinks | Arcs (000s) | Supply (000s) | FC Range | FixNetGI Best Z | Cplex Best Z | Z-Ratio | FixNetGI time (sec) | Cplex Time-X |
|---|---|---|---|---|---|---|---|---|---|---|
| 1001 | 500 | 150/350 | 10 | 100 | [20,200] | 356,689 | 355,891 | 1.002 | 2.28 | 1,582 |
| 1002 | 500 | 150/350 | 10 | 100 | [1600,6400] | 1,450,668 | 1,458,839 | 0.994 | 1.29 | 2,793 |
| 1003 | 500 | 150/350 | 10 | 500 | [20,200] | 1,615,340 | 1,614,341 | 1.001 | 3.35 | 1,075 |
| 1004 | 500 | 150/350 | 10 | 500 | [1600,6400] | 3,026,670 | 3,019,022 | 1.003 | 1.24 | 2,903 |
| 1005 | 500 | 150/350 | 50 | 100 | [20,200] | 317,018 | 317,199 | 0.999 | 14.81 | 243 |
| 1006 | 500 | 150/350 | 50 | 100 | [1600,6400] | 1,233,074 | 1,228,705 | 1.004 | 6.30 | 572 |
| 1007 | 500 | 150/350 | 50 | 500 | [20,200] | 1,519,582 | 1,519,662 | 1.000 | 16.93 | 213 |
| 1008 | 500 | 150/350 | 50 | 500 | [1600,6400] | 2,475,879 | 2,472,508 | 1.001 | 7.64 | 471 |
| 1009 | 500 | 150/350 | 100 | 100 | [20,200] | 315,383 | 315,917 | 0.998 | 16.01 | 225 |
| 1010 | 500 | 150/350 | 100 | 100 | [1600,6400] | 1,242,415 | 1,230,644 | 1.010 | 5.78 | 623 |
| 1011 | 500 | 150/350 | 100 | 500 | [20,200] | 1,515,707 | 1,516,089 | 1.000 | 16.91 | 213 |
| 1012 | 500 | 150/350 | 100 | 500 | [1600,6400] | 2,507,125 | 2,493,600 | 1.005 | 6.11 | 590 |
| 1013 | 500 | 100/100 | 10 | 100 | [20,200] | 506,218 | 505,593 | 1.001 | 3.58 | 1,006 |
| 1014 | 500 | 100/100 | 10 | 100 | [1600,6400] | 1,493,392 | 1,237,146 | 1.207 | 2.10 | 1,713 |
| 1015 | 500 | 100/100 | 10 | 500 | [20,200] | 2,417,010 | 2,416,865 | 1.000 | 2.89 | 1,245 |
| 1016 | 500 | 100/100 | 10 | 500 | [1600,6400] | 3,161,702 | 3,149,330 | 1.004 | 2.55 | 1,410 |
| 1017 | 500 | 100/100 | 50 | 100 | [20,200] | 363,544 | 362,896 | 1.002 | 9.23 | 390 |
| 1018 | 500 | 100/100 | 50 | 100 | [1600,6400] | 1,193,942 | 916,022 | 1.303 | 3.94 | 913 |
| 1019 | 500 | 100/100 | 50 | 500 | [20,200] | 1,724,593 | 1,724,192 | 1.000 | 8.79 | 410 |
| 1020 | 500 | 100/100 | 50 | 500 | [1600,6400] | 2,472,404 | 2,363,545 | 1.046 | 4.14 | 869 |
| 1021 | 500 | 100/100 | 100 | 100 | [20,200] | 344,606 | 344,442 | 1.000 | 16.84 | 214 |
| 1022 | 500 | 100/100 | 100 | 100 | [1600,6400] | 946,404 | 821,025 | 1.153 | 6.28 | 574 |
| 1023 | 500 | 100/100 | 100 | 500 | [20,200] | 1,579,353 | 1,578,955 | 1.000 | 17.13 | 210 |
| 1024 | 500 | 100/100 | 100 | 500 | [1600,6400] | 2,120,325 | 2,106,602 | 1.007 | 9.52 | 378 |
| 1025 | 1000 | 300/700 | 10 | 100 | [20,200] | 423,114 | 419,652 | 1.008 | 2.61 | 1,379 |
| 1026 | 1000 | 300/700 | 10 | 100 | [1600,6400] | 2,817,946 | 2,792,776 | 1.009 | 2.60 | 1,387 |
| 1027 | 1000 | 300/700 | 10 | 500 | [20,200] | 1,848,984 | 1,847,206 | 1.001 | 2.35 | 1,535 |
| 1028 | 1000 | 300/700 | 10 | 500 | [1600,6400] | 4,564,825 | 4,472,742 | 1.021 | 1.61 | 2,232 |
| 1029 | 1000 | 300/700 | 50 | 100 | [20,200] | 359,472 | 358,373 | 1.003 | 7.32 | 492 |
| 1030 | 1000 | 300/700 | 50 | 100 | [1600,6400] | 2,615,272 | 2,607,964 | 1.003 | 8.08 | 446 |
| 1031 | 1000 | 300/700 | 50 | 500 | [20,200] | 1,582,610 | 1,581,089 | 1.001 | 8.58 | 420 |
| 1032 | 1000 | 300/700 | 50 | 500 | [1600,6400] | 3,803,147 | 3,773,611 | 1.008 | 7.76 | 464 |
| 1033 | 1000 | 300/700 | 100 | 100 | [20,200] | 338,193 | 337,842 | 1.001 | 15.25 | 236 |
| 1034 | 1000 | 300/700 | 100 | 100 | [1600,6400] | 2,168,455 | 2,144,094 | 1.011 | 13.77 | 261 |
| 1035 | 1000 | 300/700 | 100 | 500 | [20,200] | 1,558,965 | 1,557,745 | 1.001 | 16.83 | 214 |
| 1036 | 1000 | 300/700 | 100 | 500 | [1600,6400] | 3,592,581 | 3,568,389 | 1.007 | 16.32 | 221 |
| 1037 | 1000 | 200/200 | 10 | 100 | [20,200] | 655,125 | 652,786 | 1.004 | 4.99 | 722 |
| 1038 | 1000 | 200/200 | 10 | 100 | [1600,6400] | 2,798,754 | 2,202,916 | 1.270 | 2.56 | 1,408 |
| 1039 | 1000 | 200/200 | 10 | 500 | [20,200] | 3,067,512 | 3,067,129 | 1.000 | 4.03 | 894 |
| 1040 | 1000 | 200/200 | 10 | 500 | [1600,6400] | 5,135,864 | 4,863,736 | 1.056 | 1.45 | 2,488 |
| 1041 | 1000 | 200/200 | 50 | 100 | [20,200] | 424,763 | 421,677 | 1.007 | 7.58 | 475 |
| 1042 | 1000 | 200/200 | 50 | 100 | [1600,6400] | 2,004,296 | 1,580,543 | 1.268 | 5.41 | 666 |
| 1043 | 1000 | 200/200 | 50 | 500 | [20,200] | 1,903,514 | 1,903,031 | 1.000 | 13.61 | 265 |
| 1044 | 1000 | 200/200 | 50 | 500 | [1600,6400] | 3,439,539 | 3,318,684 | 1.036 | 5.84 | 616 |
| 1045 | 1000 | 200/200 | 100 | 100 | [20,200] | 385,023 | 383,094 | 1.005 | 18.33 | 196 |
| 1046 | 1000 | 200/200 | 100 | 100 | [1600,6400] | 1,840,723 | 1,406,015 | 1.309 | 9.36 | 385 |
| 1047 | 1000 | 200/200 | 100 | 500 | [20,200] | 1,677,809 | 1,677,451 | 1.000 | 22.68 | 159 |
| 1048 | 1000 | 200/200 | 100 | 500 | [1600,6400] | 3,197,222 | 2,914,185 | 1.097 | 8.86 | 406 |

**Table 13 Test Set 2, 3000- and 5000-node problem characteristics and solution results for FixNetGI and Cplex 12.8**

| Prob | Nodes | Sources / Sinks | Arcs (000s) | Supply (000s) | FC Range | FixNetGI Best Z | Cplex Best Z | Z-Ratio | FixNetGI time (sec) | Cplex Time-X |
|---|---|---|---|---|---|---|---|---|---|---|
| 1049 | 3000 | 900/2100 | 10 | 100 | [20,200] | 659,133 | 650,375 | 1.013 | 2.76 | 1,306 |
| 1050 | 3000 | 900/2100 | 10 | 100 | [1600,6400] | 7,733,243 | 7,642,712 | 1.012 | 1.98 | 1,815 |
| 1051 | 3000 | 900/2100 | 10 | 500 | [20,200] | 2,396,668 | 2,391,344 | 1.002 | 2.39 | 1,504 |
| 1052 | 3000 | 900/2100 | 10 | 500 | [1600,6400] | 10,099,152 | 10,064,444 | 1.003 | 2.12 | 1,700 |
| 1053 | 3000 | 900/2100 | 50 | 100 | [20,200] | 498,714 | 494,887 | 1.008 | 12.13 | 297 |
| 1054 | 3000 | 900/2100 | 50 | 100 | [1600,6400] | 5,664,575 | 5,611,541 | 1.009 | 12.85 | 280 |
| 1055 | 3000 | 900/2100 | 50 | 500 | [20,200] | 1,818,914 | 1,816,890 | 1.001 | 14.22 | 253 |
| 1056 | 3000 | 900/2100 | 50 | 500 | [1600,6400] | 8,778,672 | 8,729,810 | 1.006 | 13.03 | 276 |
| 1057 | 3000 | 900/2100 | 100 | 100 | [20,200] | 455,864 | 454,198 | 1.004 | 22.16 | 162 |
| 1058 | 3000 | 900/2100 | 100 | 100 | [1600,6400] | 5,119,067 | 5,126,635 | 0.999 | 21.11 | 171 |
| 1059 | 3000 | 900/2100 | 100 | 500 | [20,200] | 1,715,184 | 1,713,425 | 1.001 | 23.56 | 153 |
| 1060 | 3000 | 900/2100 | 100 | 500 | [1600,6400] | 7,109,451 | 7,110,977 | 1.000 | 25.10 | 143 |
| 1061 | 3000 | 600/600 | 10 | 100 | [20,200] | 1,180,615 | 1,159,167 | 1.019 | 3.26 | 1,103 |
| 1062 | 3000 | 600/600 | 10 | 100 | [1600,6400] | 8,011,095 | 7,545,095 | 1.062 | 2.18 | 1,651 |
| 1063 | 3000 | 600/600 | 10 | 500 | [20,200] | 5,031,102 | 5,019,882 | 1.002 | 5.01 | 718 |
| 1064 | 3000 | 600/600 | 10 | 500 | [1600,6400] | 12,953,363 | 11,923,212 | 1.086 | 2.42 | 1,490 |
| 1065 | 3000 | 600/600 | 50 | 100 | [20,200] | 692,841 | 675,280 | 1.026 | 9.76 | 369 |
| 1066 | 3000 | 600/600 | 50 | 100 | [1600,6400] | 6,398,952 | 4,697,047 | 1.362 | 9.15 | 393 |
| 1067 | 3000 | 600/600 | 50 | 500 | [20,200] | 2,716,655 | 2,703,913 | 1.005 | 10.36 | 347 |
| 1068 | 3000 | 600/600 | 50 | 500 | [1600,6400] | 8,666,228 | 7,987,438 | 1.085 | 9.69 | 371 |
| 1069 | 3000 | 600/600 | 100 | 100 | [20,200] | 562,672 | 545,123 | 1.032 | 15.31 | 235 |
| 1070 | 3000 | 600/600 | 100 | 100 | [1600,6400] | 5,849,454 | 5,230,491 | 1.118 | 15.90 | 226 |
| 1071 | 3000 | 600/600 | 100 | 500 | [20,200] | 2,287,102 | 2,277,315 | 1.004 | 16.73 | 215 |
| 1072 | 3000 | 600/600 | 100 | 500 | [1600,6400] | 7,638,972 | 7,031,009 | 1.086 | 16.26 | 221 |
| 1073 | 5000 | 1500/3500 | 10 | 100 | [20,200] | 878,096 | 871,688 | 1.007 | 2.76 | 1,306 |
| 1074 | 5000 | 1500/3500 | 10 | 100 | [1600,6400] | 14,241,804 | 14,008,932 | 1.017 | 1.98 | 1,815 |
| 1075 | 5000 | 1500/3500 | 10 | 500 | [20,200] | 2,806,918 | 2,796,959 | 1.004 | 2.39 | 1,504 |
| 1076 | 5000 | 1500/3500 | 10 | 500 | [1600,6400] | 16,539,549 | 16,487,661 | 1.003 | 2.12 | 1,700 |
| 1077 | 5000 | 1500/3500 | 50 | 100 | [20,200] | 646,918 | 648,049 | 0.998 | 12.13 | 297 |
| 1078 | 5000 | 1500/3500 | 50 | 100 | [1600,6400] | 10,034,153 | 10,419,983 | 0.963 | 12.85 | 280 |
| 1079 | 5000 | 1500/3500 | 50 | 500 | [20,200] | 2,119,350 | 6,903,430 | 0.307 | 14.22 | 253 |
| 1080 | 5000 | 1500/3500 | 50 | 500 | [1600,6400] | 12,157,360 | 12,408,107 | 0.980 | 13.03 | 276 |
| 1081 | 5000 | 1500/3500 | 100 | 100 | [20,200] | 578,204 | 573,823 | 1.008 | 22.16 | 162 |
| 1082 | 5000 | 1500/3500 | 100 | 100 | [1600,6400] | 8,781,707 | 8,697,678 | 1.010 | 21.11 | 171 |
| 1083 | 5000 | 1500/3500 | 100 | 500 | [20,200] | 1,927,148 | 1,921,606 | 1.003 | 23.56 | 153 |
| 1084 | 5000 | 1500/3500 | 100 | 500 | [1600,6400] | 10,903,122 | 296,140,690 | 0.037 | 25.10 | 143 |
| 1085 | 5000 | 1000/1000 | 10 | 100 | [20,200] | 1,617,523 | 1,594,130 | 1.015 | 3.56 | 1,010 |
| 1086 | 5000 | 1000/1000 | 10 | 100 | [1600,6400] | 15,691,467 | 14,233,263 | 1.102 | 2.25 | 1,599 |
| 1087 | 5000 | 1000/1000 | 10 | 500 | [20,200] | 6,619,232 | 6,607,723 | 1.002 | 7.93 | 454 |
| 1088 | 5000 | 1000/1000 | 10 | 500 | [1600,6400] | 21,746,227 | 20,078,528 | 1.083 | 3.01 | 1,198 |
| 1089 | 5000 | 1000/1000 | 50 | 100 | [20,200] | 894,573 | 857,541 | 1.043 | 11.92 | 302 |
| 1090 | 5000 | 1000/1000 | 50 | 100 | [1600,6400] | 10,733,033 | 8,240,724 | 1.302 | 12.21 | 295 |
| 1091 | 5000 | 1000/1000 | 50 | 500 | [20,200] | 3,358,382 | 3,338,499 | 1.006 | 13.31 | 270 |
| 1092 | 5000 | 1000/1000 | 50 | 500 | [1600,6400] | 14,089,181 | 11,755,011 | 1.199 | 13.66 | 264 |
| 1093 | 5000 | 1000/1000 | 100 | 100 | [20,200] | 771,060 | 726,754 | 1.061 | 21.66 | 166 |
| 1094 | 5000 | 1000/1000 | 100 | 100 | [1600,6400] | 8,494,313 | 7,072,083 | 1.201 | 20.45 | 176 |
| 1095 | 5000 | 1000/1000 | 100 | 500 | [20,200] | 2,700,873 | 2,684,237 | 1.006 | 22.83 | 158 |
| 1096 | 5000 | 1000/1000 | 100 | 500 | [1600,6400] | 10,864,655 | 406,850,056 | 0.027 | 23.60 | 153 |

**Table 14 Problem group and overall average Z-ratio, FixNetGI time, Cplex Time Multiple**

| Group | Z-Ratio | FixNetGI time (sec) | Cplex Time-X |
|---|---|---|---|
| 500-node transportation | 1.001 | 8.221 | 958.4 |
| 500-node transshipment | 1.060 | 7.250 | 777.7 |
| 1000-node transportation | 1.006 | 8.589 | 773.8 |
| 1000-node transshipment | 1.088 | 8.724 | 723.3 |
| 3000-node transportation | 1.005 | 12.783 | 671.7 |
| 3000-node transshipment | 1.074 | 9.670 | 611.8 |
| 5000-node transportation | 0.861 | 12.783 | 671.7 |
| 5000-node transshipment | 1.004 | 13.033 | 503.7 |
| All | 1.012 | 10.132 | 711.5 |

Summary performance statistics by problem size and structure are given in Table 14. In terms of solution quality between the two solvers, The FixNetGI solution values average 1.2% larger than Cplex's, but for 13 of the 96 problems FixNetGI solutions are superior (Z-ratio less than 1), including some larger instances where Cplex's Best Z is 30 times larger. Based on average Z-ratio, the heuristic's solution quality tends to be superior for transportation problems when compared to transshipment problems with the same number of nodes.

In terms of solution speed, Cplex runs to the one-hour time limit in all cases. FixNetGI averages 10.1 seconds per problem, or 700 times faster than Cplex's 3600-seconds, as shown in the Cplex Time-X column of Table 14. These multiples are better for the smaller problems, but all multiples would be much larger if Cplex had been allowed to run to optimality.

## Test Set 2: Computational Results and Statistical Analysis

The structure of the test set enables rigorous statistical analysis of the relative performance of Cplex and FixNetGI solvers in terms of solution values and solution time, and the effect of the five factors described above. SAS 9.2's analysis of variance procedure (ANOVA) and comparisons of means using Tukey's Significant Difference (TSD) test are employed to determine whether the average results differed by solution method and whether factors affected the average results. The TSD procedure compares and ranks solver performance under the effect of different single-factor levels and treatment combinations. Specifically, we test hypotheses that the mean solution times and solution values are the same for both solvers and under different factor levels.

Based on the problem solution times and values in Tables 12 and 13, ANOVA shows a statistically significant difference in mean solution times between the Cplex and FixNetGI codes. Hence, as expected, the mean solution speeds of the two solvers are statistically different, with FixNetGI being the faster. Statistical differences in time are also found between the four levels of problem node count, the two fixed-charge ranges, transportation and transshipment network structures, the three levels of number of problem arcs, and two levels of total supply and demand. Hence, all hypotheses of equivalent means are rejected when runtime is the performance metric.

However, when comparing solvers based on problem solution values (Z), the TSD test finds no statistically significant difference between the solvers. Therefore, while the mean Z-ratio for FixNetGI is slightly higher than Cplex's, ANOVA shows that the mean solution values are not statistically different and the hypothesis of equality of mean solution values is not rejected. The two fixed-charge ranges do produce statistically different average solution values, as expected, but transportation and transshipment problems do not demonstrate statistically different values, nor do the numbers of problem arcs. Problems with 5000 nodes had mean solution values that are statistically different from those with 500 and 1000 nodes, but not those with 3000 nodes.

This combination of hypothesis outcomes validates the effectiveness and speed of the GI/TS algorithm as implemented in FixNetGI for these larger and more challenging problem types. With solution times three orders of magnitude faster than Cplex while producing comparable objective function values, this approach advances the state of the art for fixed-charge network problems and renders solvable large practical instances from industrial settings.

## 5. Conclusions

Statistical testing reveals that the FixNetGI code is not only dramatically faster than Cplex in identifying its best solutions, but its mean solution quality is statistically equivalent to that of Cplex. This implementation of the GI/TS algorithm makes it appropriate for applications requiring high-quality results quickly, as in time-critical logistics, military response, airline re-scheduling, telecommunications and content-delivery network reconfiguration for demand fluctuations, and other near-real-time decision-making situations.

There are a variety of opportunities to improve the GI/TS algorithm in the future. The tabu search procedure currently employed in the method is exceedingly simple, and a more advanced version may well enhance overall performance. Another conspicuous opportunity for future improvement will be to determine better parameters settings (for example, based on problem size and network class). A related possibility for investigation is to shortcut the Inside Loop operation and solve LP(p) more often, with the option of updating the solution each time by solving the restricted LP problem. Within the DUPCHECK procedure, the trade-offs between the sLim and the LimMatch values likewise invite examination, as do the values of the "alpha parameters" in V_UPDATE.

The attractive outcomes produced by the current version of GI/TS embodied in FixNetGI provides a significant advance in our ability to solve fixed cost network problems efficiently and motivates a study devoted to the solution of practical problems in multiple areas.